\documentclass[]{jfm}

\usepackage{graphicx}
\usepackage{newtxtext}
\usepackage{newtxmath}
\usepackage{natbib}
\usepackage{hyperref}
\hypersetup{
    colorlinks = true,
    urlcolor   = blue,
    citecolor  = black,
}

\newcommand{\RomanNumeralCaps}[1]

\def \NC#1 {\textcolor{red}{#1} }

\usepackage{color}
\usepackage{float}
\usepackage{caption}
\usepackage{subcaption}

\title{Long-time evolution of density layers and interfaces in forced stably-stratified flows}

\author{Niccolò Cocciaglia\aff{1}
\corresp{\email{\textit{niccolo.cocciaglia@roma2.infn.it}}},
Fabio Bonaccorso\aff{1}, 
Alessandra Sabina Lanotte\aff{2}
\and Luca Biferale\aff{1}
}

\affiliation{\aff{1}Department of Physics \& INFN, University of Rome ``Tor Vergata'',
Via della Ricerca Scientifica 1, 00133, Rome, Italy
\aff{2}CNR NANOTEC \& INFN Sezione di Lecce, Via Monteroni, 73100 Lecce, Italy}

\begin{document}
\maketitle

\begin{abstract}

Stably stratified fluids subject to sustained forcing are known to develop step-like density “staircases”, where nearly homogeneous layers alternate with thin interfaces of strong stratification. 
However, long-time numerical investigations of this phenomenon have been limited by the intrinsically slow evolution of large-scale modes and the sensitivity of stratified turbulence to physical parameters. 
We present direct numerical simulations of forced Boussinesq flows for three stratification strengths ($Fr=0.42, 0.22, 0.076$) and of unprecedented time extensions -- up to $O(10^4)$ turnover times -- with the purpose of reproducing and studying the very slow coarsening of the layered state. 
A large-scale friction term is introduced to arrest shear-mode growth and mimic finite-domain constraints. 
Staircase formation is observed for both medium and strong stratified cases, following two different coarsening dynamics: interfaces decaying or merging. While kinetic energy remains quasi-stationary during interface decay, it exhibits sharp bursts during merging events. 
The emergence and persistence of density steps can be explained by the non-monotonic relation between buoyancy flux and buoyancy gradient. 
Intermittency in vertical velocity and density fluctuations is confined to the vicinity of layer–interface boundaries, indicating that strong events arise from the interaction between turbulent mixing and layer formation rather than from regions of large density gradients alone.
\end{abstract}

\section{Introduction}
\label{sec:intro}
The large-scale and long-time behavior of oceanic and atmospheric flows is greatly influenced by Earth's rotation and density variations \citep{gill_book, davidson_book}. 
The visible effect is the formation of highly anisotropic structures: the density stratification under gravity hinders motions in the vertical direction so that flattened layers, often called ``pancake structures'' \citep{billant2000experimental}, form throughout the fluid, whereas the Coriolis acceleration tends to generate rotating columnar vortices \citep{davidson_book}. 
Nontrivial effects stemming from the nonlinear interaction between waves, vortices and small-scale turbulence \citep{lelong1991internal, bartello1995} make the evolution of rotating stratified turbulence very hard to predict and model. 

A characteristic phenomenon of stratified, non-rotating flows is the development of thin regions with a large vertical density gradient, called henceforth ``density interfaces'', enclosed between ``density layers'' characterized by nearly-homogeneous density: this can be visualized in figure~\ref{fig:visual_layers}, panels (c-f,h). 
We refer to \citet{caulfield2021review} for a recent review on the subject. 
Much effort has been put into understanding and reproducing the splitting between well-mixed and high-gradient regions, and their subsequent evolution in time. 
\citet{orlanski1969formation}, following ideas brought forward by \citet{munk1966abyssal}, claimed that overturning events caused by shear or convective instabilities of internal-gravity waves are the main source of mixing in ocean layers. 
Experimental investigations followed, in which a stably-stratified fluid 
was stirred using either horizontal grids oscillating vertically around their mean position (\citet{linden1979mixing} and references therein) or arrays of vertical rods moving along the length of the tank \citep{ruddick1989formation}.
Both stirring mechanisms indeed generate sharp vertical gradients separated by layers with approximately uniform density (i.e. salt concentration).
\citet{park1994turbulent} tracked with detail the evolution of density interfaces, and discriminated between layered and non-layered cases in terms of energetic considerations and mixing efficiency. 

The energetics of a stratified turbulent flow, with or without background rotation, follows a more complicated behavior than that of Homogeneous Isotropic Turbulence (HIT). 
In HIT the statistical balance between energy injection and dissipation, mediated by the energy cascade, generates a persistent stationary state. 
On the other hand plateaus of kinetic (and potential) energy, often found in stratified turbulence \citep{lindborg2006cascade, howland2020mixing}, are of a metastable and transient nature, even if very long-lived.
A clear manifestation of such transient states is the evolution in time of the layered density structure we described earlier, whose analysis is the focus of the present study. 

Motivated by experimental results, attempts have been made at modeling the dynamics of density layers and interfaces. 
\citet{balmforth1998dynamics} proposed and employed a reduced model of mixing in stratified flows based only on the horizontally-averaged vertical density gradient and the kinetic energy density. 
The numerical solutions displayed both merging and decaying of the interfaces, a very nontrivial phenomenon already found in the experiment by \citet{park1994turbulent}.
Merging of interfaces occurs when two flat, high-gradient regions move vertically toward each other, merge, and form a single steeper interface, whereas the decay of an isolated interface, arguably due to instabilities leading to mixing, has the effect of removing the ``barrier'' between adjacent layers and generating a thick region with homogeneous density. 
\citet{radko2007mechanics} performed a linear stability analysis on an initial step-like vertical profile of the density, and a criterion for the preferred mechanism of coarsening -- i.e. reduction of the number of interfaces -- was devised. 
It was confirmed by numerical tests that both merging and decaying may be observed in the Balmforth \textit{et al.} model at varying initial conditions, while a model of double diffusion by \citet{merryfield2000origin} displayed only decaying as confirmed by Radko's analysis. 

While investigated in reduced models of buoyancy-driven flows, the late-time evolution of layers and interfaces after their appearance (approximately the first-observed metastable state) was not addressed by numerical studies of the full Boussinesq equations, probably due to the considerable amount of time necessary for the development of such structures. 
Even though buoyancy forces bring the system to a highly-anisotropic state that looks almost one-dimensional (at late-enough times little horizontal variability is noticeable), it is nonetheless important to track the evolution of the whole velocity and density fields to understand better complex spatial interactions and avoid criticisms connected to the use of reduced models.
In order to extend simulations for very long times the spatial resolution needs to be quite modest, so small-scale effects may not be well-reproduced. 
Although interactions among widely different scales are important in stratified flows, especially in wave-breaking events \citep{muller1986nonlinear}, it may be conjectured that the coarsening process of density interfaces is not determined by spectral interactions with dissipative scales, thus unresolved spectral components should not influence the observed phenomenology.

In this work we investigate the long-time behavior of stratified fluids forced isotropically at large scales, for three values of the buoyancy strength, with a focus on the emergence and subsequent space-time evolution of density layers and interfaces. 
In section~\ref{sec:eqs_and_numerics} we specify the equations and numerical details of our simulations. 
The evolution of both kinetic and potential energies is treated in section~\ref{sec:energy_evolution}, where we point out how metastable states and non-stationary transient periods are associated with the re-adjustment of the vertical density profile, mediated or not by strong bursts of velocity. A spectral analysis is also present. 
In section~\ref{sec:flux_gradient} we analyze the relationship between the vertical flux and the vertical gradient of the buoyancy (a properly-rescaled density) in the spirit of the instability theory by Phillips and Posmentier.
Section~\ref{sec:decay_merge} is dedicated to the processes of interfaces decay and merging, that determine the coarsening of the vertical profile, while in section~\ref{sec:intermittency} the statistics of rare events is characterized, highlighting the differences between where the density profile is flat and where it is instead steep. 
We end with conclusions and some perspectives in section~\ref{sec:conclusion}.


\section{Equations and Numerical Details}
\label{sec:eqs_and_numerics}
In the Boussinesq approximation, the density fluctuations around the reference stably-stratified state are modeled as an active scalar $\phi(\boldsymbol{r})$ advected by the velocity field $\boldsymbol{u}(\boldsymbol{r})$. 
The fluid density $\rho(\boldsymbol{r})$ has a linear vertical profile, decreasing with height, plus small perturbations superimposed to it: 
\begin{equation}
    \rho(\boldsymbol{r}) = \tilde{\rho}(z) +\delta \rho(\boldsymbol{r}) = \rho_0 -\sigma z +\delta \rho(\boldsymbol{r})\ ,
    \label{eq:rho_decomp}
\end{equation} 
where $\tilde{\rho}(z)$ is the unperturbed density profile, $\rho_0$ the ambient density and $\sigma = \left\rvert d\tilde{\rho}/dz \right \rvert$ is the modulus of the (constant) density gradient parallel to gravity $\boldsymbol{g} = -g\hat{z}$. 
The relationship between the scalar field (with physical dimensions of a velocity) and the density fluctuation $\delta \rho(\boldsymbol{r})$ is:
\begin{equation}
    \phi(\boldsymbol{r}) = \frac{g}{N \rho_0}\, \delta \rho(\boldsymbol{r})\, ,
    \label{eq:phi_deltarho_relation}
\end{equation}
where $N = \sqrt{g \sigma/\rho_0}$ is the Brunt-Vais\"al\"a frequency. 

The Boussinesq equations coupling scalar and velocity fields read: 
\begin{align}
    \partial_t \boldsymbol{u} + (\boldsymbol{u}\cdot \boldsymbol{\nabla}) \boldsymbol{u} &= -N \phi \hat{z} - \boldsymbol{\nabla}P/\rho_0 - \nu_q (-\Delta)^q \boldsymbol{u} +L(\boldsymbol{u}) + \boldsymbol{f}\ , \label{eq:u_evol} \\[.1cm]
    \partial_t \phi + (\boldsymbol{u} \cdot \boldsymbol{\nabla}) \phi &= N \boldsymbol{u} \cdot \hat{z} - \kappa_q (-\Delta)^q \phi \ . \label{eq:phi_evol}
\end{align}
Hyperdissipation and hyperdiffusion of order $q=2$ are used, with coefficients $\nu_q$ and $\kappa_q$ respectively, and a Fourier-filtered large-scale damping term $L(\boldsymbol{u}) = \left[\alpha\, \Delta^{-1} \boldsymbol{u}\right]\big\rvert_{k_{min} \leq k \leq k_{max}}$, active only inside the  Fourier shell between $k_{min}=0.5$ and $k_{max}=2.5$, constrains the large-scale velocity. 
A stronger-than-linear damping is chosen (inverse Laplacian) to mimick wall friction and arrest the unbounded growth of shear modes with $k_{\perp}\equiv (k_x^2+k_y^2)^{1/2}=0$ \citep{lindborg2006cascade}. 
Parallel and perpendicular directions are intended with respect to the gravity acceleration direction $\hat{z}$.
No explicit forcing on the scalar field is present, so it receives energy only through the exchange term $N \boldsymbol{u} \cdot \hat{z}$. 

The equations are integrated numerically with a pseudo-spectral code in a cubic domain of length $M=2\pi$, using 2nd-order Adams-Bashforth time stepping and $2/3$ dealiasing rule. 
The Ornstein-Uhlenbeck-type forcing \citep{sawford1991reynolds} is restricted to a spectral shell with $4\leq |\boldsymbol{k}|\leq 6$. 
The forcing is isotropic, thus both internal waves and horizontal motions are directly excited. 
Multiple simulations have been performed varying the stratification strength ($N=3,6,18$) to point out three different phenomenologies of interfaces evolution. 
In order to achieve long integration times -- up to $O(10^4)$ turnover times -- a coarse spatial domain of $256^3$ collocation points has been chosen. 
The parameters shared by all runs are the ones written in table~\ref{tab:parameters}. 
Other useful quantities, computed from parameters or from measurements during the runs, are listed in table~\ref{tab:scales_derivedparameters}. 
\begin{table}
  \begin{center}
  \def~{\hphantom{0}}
  \begin{tabular}{ccc}
      Parameter & Symbol &  Value \\[3pt]
      Linear grid resolution & $n$ & $256$ \\
      Time step & $dt$ & $5\times10^{-4}$ \\
      Domain length & $L$ & $2\pi$ \\
      Hyper-viscosity & $\nu_q$ & $1.6 \times 10^{-6}$ \\
      Hyper-diffusity & $\kappa_q$ & $1.6 \times 10^{-6}$ \\
      Friction coefficient & $\alpha$ & $0.1$ \\
      Forcing corr. time & $\tau_f$ & $0.023$ \\ 
      Grid spacing & $dx$ & 0.0245 \\
  \end{tabular}
  \caption{Values of the parameters used in numerical simulations.}
  \label{tab:parameters}
  \end{center}
\end{table}
\begin{table}
  \begin{center}
  \def~{\hphantom{0}}
  \begin{tabular}{ccccccccccc}
      $N$ & $L_f$ & $U_\text{rms}$ & $\tau_\text{NL}$ & $\varepsilon_{in}$ & $\varepsilon_K$ & $\varepsilon_P$ & $\eta_q$ & $l_\text{Oz}$ & $\text{Re}_q (\times 10^6)$ & Fr \\[3pt]
      \hline
      $3$ & $2\pi/5$ & 1.6 & 0.79 & 1.5 & 0.91 & 0.014 & 0.018 & 0.24 & 2.0 & 0.42 \\ 
      $6$ & $2\pi/5$ & 1.7 & 0.76 & 1.5 & 1.0 & 0.013 & 0.018 & 0.081 & 2.1 & 0.22 \\ 
      $18$ & $2\pi/5$ & 1.7 & 0.73 & 1.3 & 0.99 & 0.018 & 0.018 & 0.015 & 2.1 & 0.076 \\
  \end{tabular}
  \caption{Characteristic scales and non-dimensional parameters computed for different runs (different $N$). Time-dependent quantities, like $U_\text{rms}$, $\varepsilon_K$ and so on, change (even if slightly) as the layering dynamics evolves: their reported value is the time-average on the last-measured stationary state.}
  \label{tab:scales_derivedparameters}
  \end{center}
\end{table}
They include the forcing scale $L_f=2\pi/k_f$, where $k_f=5$ is the centroid of the forced shell
, 
the r.m.s. velocity $U_\text{rms}$, 
the non-linear eddy turnover time $\tau_{NL} = L_f / U_\text{rms}$, 
the kinetic energy input rate $\varepsilon_{in} = \langle \boldsymbol{f}\cdot \boldsymbol{u} \rangle$ (with angle brackets indicating here a space-time average), the kinetic and potential energy dissipation rates, respectively $\varepsilon_K = \nu_q \langle |\boldsymbol{\nabla}^q \boldsymbol{u}|^2 \rangle$ and $\varepsilon_P = \kappa_q \langle |\boldsymbol{\nabla}^q \phi|^2 \rangle$. 
The hyper-viscous Kolmogorov scale and the Ozmidov scales are computed as $\eta_q=\left(\nu_q^3/\varepsilon_K\right)^{1/(6q-2)}$ and $l_\text{Oz}=(\varepsilon_{in}/N^3)^{1/2}$.
Finally, the hyper-viscous Reynolds number $\text{Re}_q=U_\text{rms} L_f^{2q-1}/\nu_q$ and the Froude number $\text{Fr}=U_\text{rms}/(L_f\, N)$ are included. 
While we are aware that quantities involving hyper-dissipative parameters convey limited physical information (notice the extreme values of $Re_q$), we computed and reported them for completeness sake, knowing that comparisons with dissipative parameters ($q=1$) are to be done with caution.
We remark that, as we will see in figure~\ref{fig:kin_pot_energies}(a), the kinetic energy is mostly stationary while the potential energy presents a peculiar non-decreasing behaviour, so buoyancy-dependent quantities will change in unison with the scalar field and are thus more indicative of the layer formation mechanism.

The starting configuration at time $t=0$ is a linearly-stratified fluid at rest, namely both $\boldsymbol{u}$ and $\phi$ are zero everywhere. 
The energy injected into the velocity field will redistribute from the narrow wavenumber band to all other scales, and via the exchange term the velocity will excite density fluctuations. 
We end this section by clarifying the notation used in the rest of the article for indicating averages: an overbar $\overline{(\cdot)}$ refers to a time-average, whereas angle brackets without footing $\langle (\cdot) \rangle$ represent an average on the $xy$-plane.


\section{Energy evolution and connection with layering}
\label{sec:energy_evolution}

\begin{figure}
    \centering
    \includegraphics[width=\linewidth]{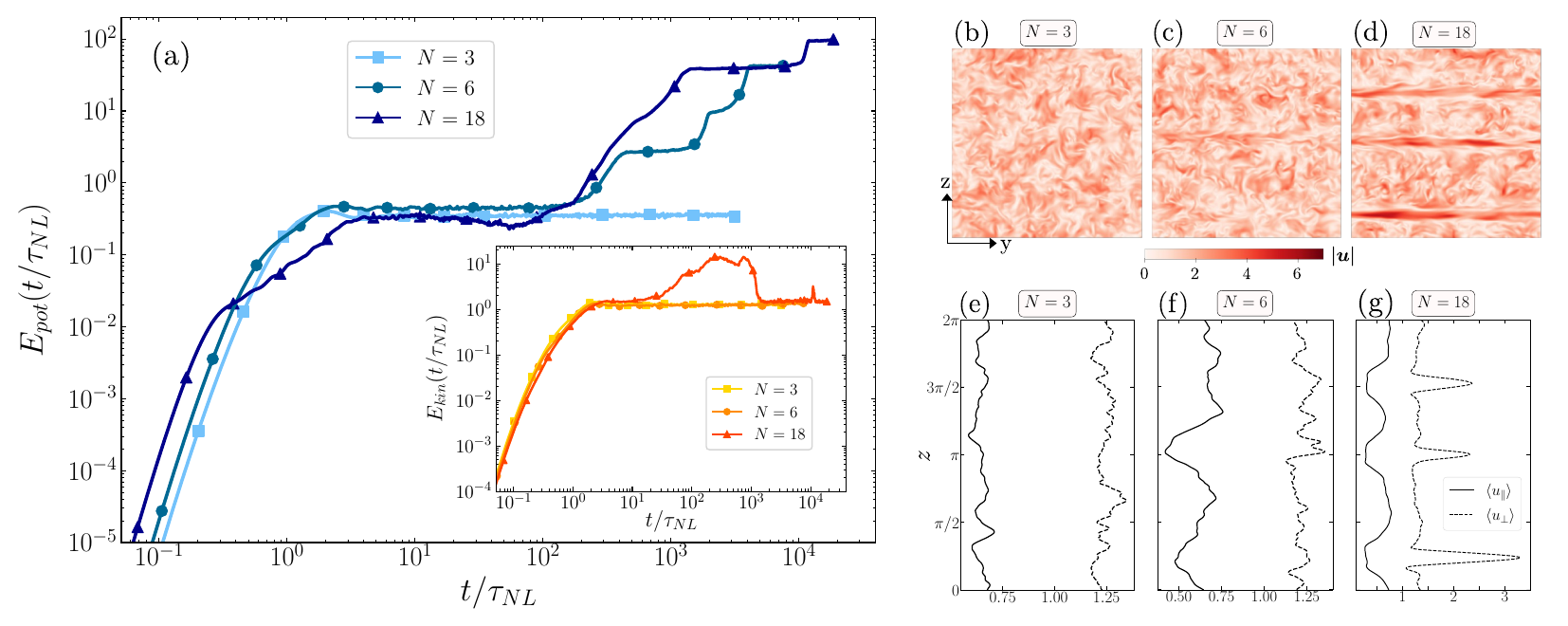}
    \caption{Evolution of kinetic and potential energies. 
    (a) Log-log plots of the time evolution of potential (main panel) and kinetic (inset) energies, for three values of the Brunt-Vais\"al\"a frequency. Time is rescaled with the large-eddy turnover time $\tau_{NL}$. 
    (b-d) Vertical $yz$ slices of the velocity magnitude field, for increasing values of $N$ at $t/\tau_{NL} \simeq 2800$. 
    (e-g) Vertical profiles of the parallel and perpendicular velocities for increasing $N$, computed at the same instant as (b-d).}
    \label{fig:kin_pot_energies}
\end{figure}

We start by focusing on the evolution of the mean kinetic and potential energies (per unit mass) 
\begin{equation}
    E_{kin}(t) = \frac{1}{2}\langle |\boldsymbol{u}(\boldsymbol{r},t)|^2\rangle_V\ , \qquad E_{pot}(t) = \frac{1}{2} \langle \phi^2(\boldsymbol{r},t) \rangle_V\ ,
    \label{eq:enekin_enepot}
\end{equation}
with $\langle (\cdot) \rangle_V$ denoting the volume average, at varying the stratification strength. 
The main panel of figure~\ref{fig:kin_pot_energies}(a) shows the increase of potential energy for the three cases. 
After an initial transient lasting $\sim 10$ large-eddy turnover times the depicted energies seem to have reached stationarity at similar values. 
However only the fluid with weaker stratification ($N=3$) remains in that steady state, while in the two cases with larger $N$, after $O(10^2)$ turnover times, the potential energy undergoes a sequence of monotonic increases followed by stationary periods. 
We identify the periods in which the potential energy is stationary, but are followed by sharp build-ups, as metastable states. 
Looking instead at the kinetic energies (figure~\ref{fig:kin_pot_energies}(a), inset), the initial transients are followed by long-lasting stationary states in the simulations with $N=3$ and $N=6$, while for the strongest stratification ($N=18$) two distinct energy excursions are evident -- the second being less intense and appearing narrower due to the logarithmic time scale -- in the form of build-ups followed by decays to their previous energy. 
These excursions for $N=18$ occur simultaneously to the two monotonic increases of potential energy. 
Panels (b-d) represent vertical $yz$ slices of the velocity magnitude for increasing stratification strength, from $N=3$ to $N=18$. 
They are shown at $t/\tau_{NL} \simeq 2800$, when potential energies are approximately stationary and the kinetic excursion for $N=18$ have decayed to the stationary value.
One sees, as $N$ grows, the emergence of thin horizontal slabs where horizontal velocity is magnified and the vertical one is further suppressed. 
This is evident in figures~\ref{fig:kin_pot_energies}(e-g) displaying the vertical profiles of the parallel (i.e. vertical) velocity $u_\parallel = |u_z|$ and the perpendicular (i.e. horizontal) velocity $u_\perp = \sqrt{u_x^2+u_y^2}$.
Vertical profiles are obtained by averaging the quantities of interest on $xy$ planes.

Numerical simulations \citep{smith2002generation} have shown that non-rotating stratified flows manifest a slow accumulation of kinetic energy at large horizontal scales ($k_\perp=0$), in the form of vertically-sheared horizontal flows or VSHF modes. 
The resulting monotonic increase of the total kinetic energy hinders the approach to the desired steady state. 
This phenomenon is usually avoided by restricting the study before the energy of the shear modes starts dominating \citep{waite2006waves}, and in several simulations with limited time duration the growth has not started yet, and the problem not posed. 
\citet{waite2004vortical} also claimed that the energy accumulation may be due to the absence of boundaries, an inevitable condition in pseudo-spectral simulations. 
For this reason, as did by \citet{lam2021energy}, a damping coefficient has been introduced in~\eqref{eq:u_evol} to reproduce large-scale friction, a common procedure when inverse cascades and condensate formation generate system-size structures, e.g. in 2D turbulence \citep{boffetta2012twod}, rotating flows \citep{alexakis2018cascades} but also rotating convection \citep{dewit2025}.
Following \citet{lam2021energy}, a damping only on the horizontal velocity would have sufficed, since the vertical velocity has much smaller magnitude. For consistency with the isotropic forcing, and considering the damping along the vertical to be subdominant, a 3D large-scale dissipation has been chosen instead.

The surges in potential energy and the occasional excursions in kinetic energy in the run with the strongest stratification are consequences of the anisotropic spatial redistribution of the density field. 
As observed in laboratory experiments (\citet{park1994turbulent} and references therein) a stable, linear density profile develops convective and shear instabilities when subject to an external stirring. 
Overturnings create localized mixed regions that will spread horizontally until homogeneous density layers will form \citep{davidson_book}. 
These are less dense on the top and heavier on the bottom due to gravitational stability, and very thin interfaces displaying large vertical density gradient connect adjacent layers, resulting in an overall staircase profile for the density. 
\begin{figure}
    \centering
    \includegraphics[width=\linewidth]{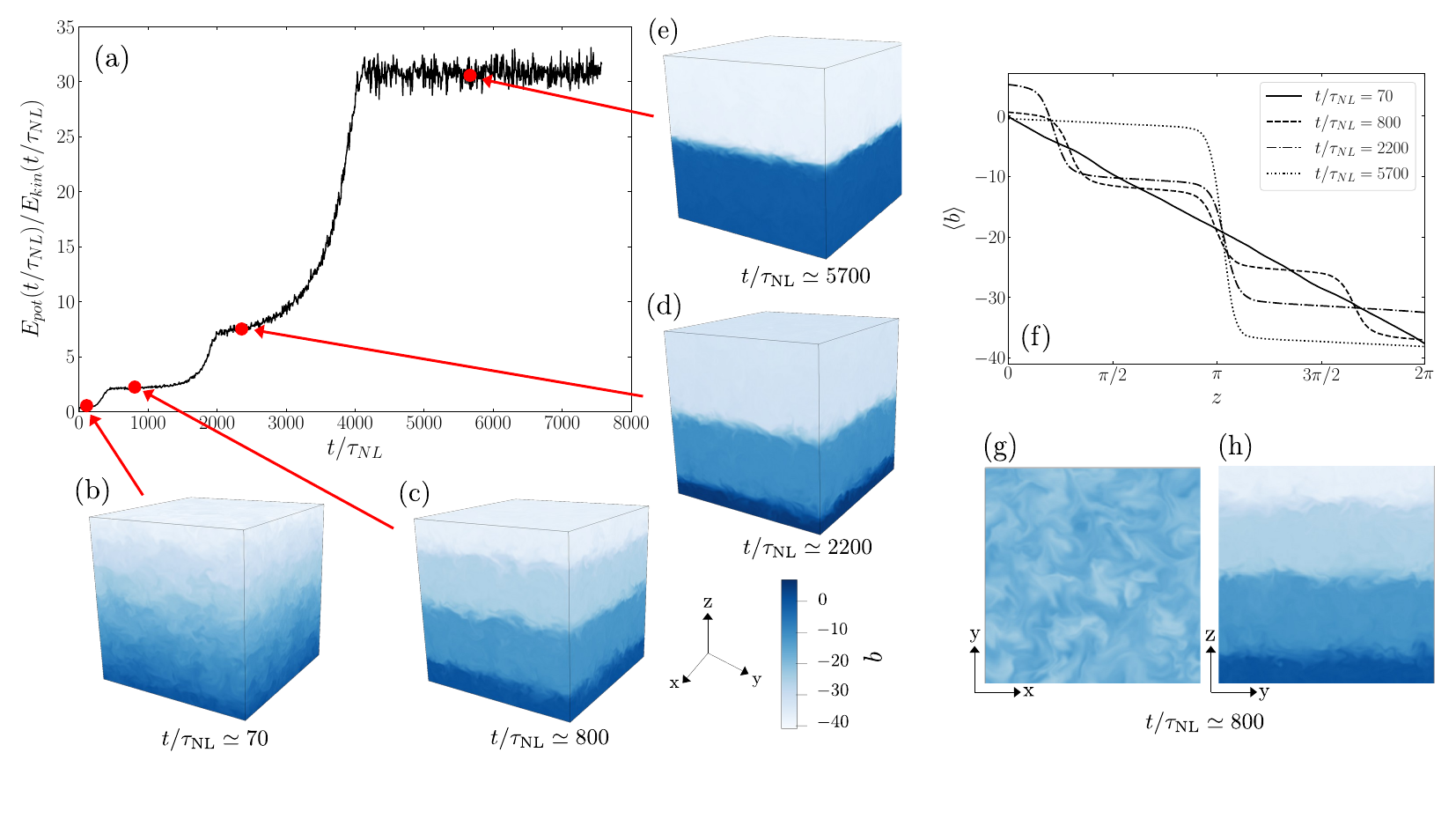}
    \caption{Visualizations of the buoyancy field $b(\boldsymbol{r},t)$ at different stages of the evolution for the $N=6$ case. 
    (a) Evolution of potential-to-kinetic-energy ratio. Along the curve four instants are chosen in which snapshots of the buoyancy $b$ are visualized. 
    (b) Buoyancy field at (comparatively) small times, displaying a moderately-perturbed linear variation along the vertical. 
    (c) First energy plateau, with buoyancy (read density) arranged as homogeneous layers separated by three density interfaces. 
    (d) Decay of an interface when the next plateau is reached. 
    (e) Final stationary state, with only one interface left.
    (f) Vertical profiles of the buoyancy at the four instants of (b-e). 
    (g) Horizontal slice of (c) taken at the height of the `middle' interface. 
    (h) Vertical slice on the $yz$ plane.}
    \label{fig:visual_layers}
\end{figure}
In figure~\ref{fig:visual_layers}(b-e) we show, for the $N=6$ run, the density field  at different instants as tracked in figure~\ref{fig:visual_layers}(a) by the changes of the potential-to-kinetic energy ratio. 
The quantity of interest here is the variable part of the density field rescaled to have dimension of velocity, i.e. the buoyancy:
\begin{equation}
    b(\boldsymbol{r},t) \equiv \frac{g}{N}\frac{\rho(\boldsymbol{r},t)-\rho_0}{\rho_0} = -Nz + \phi(\boldsymbol{r},t)\ ,
    \label{eq:buoyancy}
\end{equation} 
but the terms `density' and `buoyancy' will be often used interchangeably.
At the initial stages (panel (b)) the density still looks linearly stratified, with superimposed fluctuations and some convectively-unstable events. At the first visible energy plateau (panel (c)) the splitting of the fluid into distinct homogeneous layers, separated by three density interfaces, has taken place. 
Each successive (meta)stable state, visualized in panels (d) and (e) and occurring after a rise in potential energy, results from the decay of one density interface with negligible vertical displacement of the surviving ones. 
In fig.~\ref{fig:visual_layers}(f) we show the vertical density profiles of the four 3D snapshots, where the formation and modification of the staircase is appreciated. 
Interfaces separating two layers have a very smooth vertical profile, yet here there is considerable motion induced by internal gravity waves, as observed in horizontal and vertical cuts in figures~\ref{fig:visual_layers}(g) and \ref{fig:visual_layers}(h). 
Although the stochastic large-scale forcing breaks wave patterns, some wave fronts among the isotropic background are visible in the horizontal slice taken inside an interface, and wave-breaking events are seen as well in the vertical slice.

An interesting connection is that the staircase vertical profile for $\rho$ corresponds to a ``ramp-cliff'' structure for the $\phi$ profile, as can be appreciated in the top plots in figure~\ref{fig:vertical_profiles}. The latter has been observed for passive scalar advection, either in experiments with shear flows \citep{gibson1977structure} or in simulations with an imposed mean gradient on the scalar \citep{pumir1994numerical}.

The case $N=3$ with weaker stratification displays no layering: the density field remains indefinitely as in figure~\ref{fig:visual_layers}(b). 
On the other hand, we see from figure~\ref{fig:kin_pot_energies} that for $N=18$ the potential energy passes through stationary periods alternated to monotonic increases, similarly to $N=6$. 
Indeed density layers and interfaces emerge also here, but the evolution mechanism leading to the decreasing of the number of layers is much different, as will be discussed in section~\ref{sec:decay_merge}. 

\begin{figure}
    \centering
    \includegraphics[width=\linewidth]{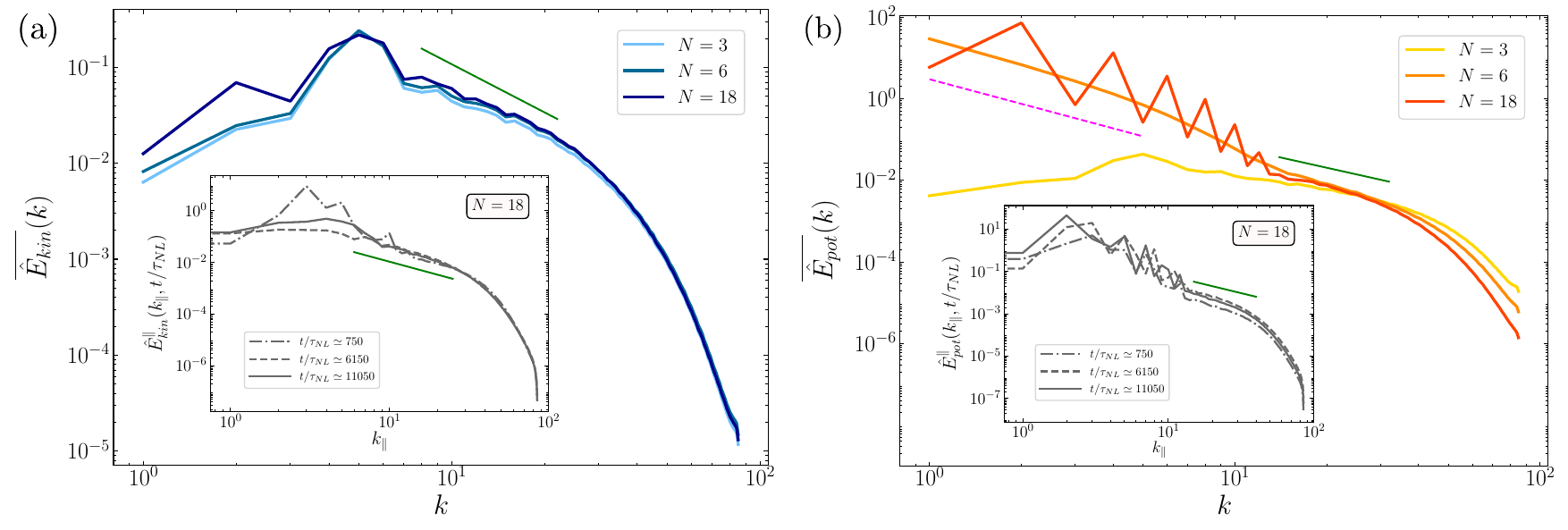}
    \caption{Kinetic and potential energy spectra. (a) Main: isotropic kinetic spectra, for all stratification strengths, time-averaged at the last stationary state reached. 
    Inset: instantaneous, parallel kinetic spectra computed for $N=18$ at three salient times: first kinetic excursion with corresponding potential energy increase (dash-dotted); following stationary state (dashed); second, minor kinetic surge with simultaneous potential energy growth (solid). 
    (b) Same as (a), but for potential energy spectra. 
    Green solid lines indicate the spectral slope $k^{-5/3}$, the magenta dashed line corresponds to $k^{-2}$.
    Time averages in main panels are taken in an interval of length $t/\tau_{NL} \simeq 650$. A short time-average (5 consecutive samples) is also performed in the inset plots, to decrease fluctuations.}
    \label{fig:spectra}
\end{figure}

To understand the effects of layering on the spectral redistribution of energy we show in figure~\ref{fig:spectra} both kinetic and potential energy spectra. Main panels show the typical isotropic spectra
\begin{equation}
    \hat{E}_{kin}(k,t) = \frac{1}{2} \sum_{k\leq |\boldsymbol{k}| < k+1} |\boldsymbol{u}(\boldsymbol{k},t)|^2\ , \qquad \hat{E}_{pot}(k,t) = \frac{1}{2} \sum_{k\leq |\boldsymbol{k}| < k+1} \phi^2(\boldsymbol{k},t)\ ,
    \label{eq:spectra_iso}
\end{equation}
with a time-averaging performed on the last-reached stationary state. 
Insets show instead the (instantaneous) parallel spectra
\begin{equation}
    \hat{E}^\parallel_{kin}(k_\parallel,t) = \frac{1}{2} \sum_{k_\parallel \leq |k_z| < k_\parallel+1} |\boldsymbol{u}(\boldsymbol{k},t)|^2\ , \qquad \hat{E}^\parallel_{pot}(k_\parallel,t) = \frac{1}{2} \sum_{k_\parallel \leq |k_z| < k_\parallel+1} \phi^2(\boldsymbol{k},t)\ .
    \label{eq:spectra_para}
\end{equation}
Kinetic stationary spectra in the main panel of figure~\ref{fig:spectra}(a) are peaked at the forced scales, and for $N=18$ a large-scale accumulation of kinetic energy is evident with respect to the other cases. 
Regarding this aspect, parallel kinetic spectra in the inset of figure~\ref{fig:kin_pot_energies}(a) reveal that during the kinetic energy excursions the growth is concentrated at or below the forced vertical scales. 
The Kolmogorov-like scaling law expected in the weak stratification regime, namely $\hat{E}_{kin}(k) \sim \hat{E}_{pot}(k) \sim k^{-5/3}$ (represented by green solid lines), reproduces well the slope of the parallel kinetic spectra, whereas the isotropic time-averaged ones appear shallower. 
In figure~\ref{fig:spectra}(b), main panel, the $N=3$ case is immediately recognizable by the absence of potential energy accumulation below the forced scale, i.e. the absence of layering. 
While for $N=6$ we have reached the ``final'' stable state with one interface and the spectral energy peaked at $k=1$, for $N=18$ we arrived at a two-interfaces state -- as will be evident from figure~\ref{fig:decay_merge}(b) -- thus the $k=2$ mode and its multiples store most of the spectral energy.
The latter simulation has not been continued to reach the one-interface state and thus obtain the smooth spectrum dominated by the lowest non-zero mode as shown by $N=6$ (more comments on this in section~\ref{sec:decay_merge}). 
The large-scale $k^{-2}$ scaling displayed in the potential energy spectrum has been observed by \cite{waite2011stratified}, and generally associated to a 2-dimensionalization of the system at the mesoscales.
The change of regime (and scaling) occurring for $N\neq 3$ at $k\simeq12$ do not seem to be dictated by the buoyancy scale $k_b = N/U_{rms}$ -- which should be evidently different in the two runs -- but rather by the typical width of the interfaces that form in the fluid.
The small-scale potential energy depletes as $N$ increases, a phenomenon that is not observed for the kinetic energy. 
This is consistent with previous studies \citep{maffioli2016mixing, feraco2018vertical} showing that $\Gamma=\varepsilon_P/\varepsilon_K$, called either mixing coefficient or irreversible mixing efficiency, increases at increasing $Fr$ in the parametric range of our simulations.
From the parallel potential spectra for $N=18$ in the inset one notices the progressive shift of the energy-dominating modes towards lower $k_\parallel$.
In general, despite order-2 hyperviscosity and hyperdiffusion the inertial ranges of both velocity and scalar spectra have limited extension: using higher-order dissipative terms, for instance order-8 as in \cite{smith2002generation} or \cite{kurien2014effect}, would have extended the self-similar range of scales.


\section{Instabilities and the relation between buoyancy flux and buoyancy gradient}
\label{sec:flux_gradient}

In a stirred stratified fluid the initial development of a layered, step-like profile starting from an equilibrium linear density profile is commonly explained with the \cite{phillips1972} and \cite{posmentier1977} mechanism. 
Let us consider again the buoyancy \eqref{eq:buoyancy}, evolving according to an advection-diffusion equation (upon substituting \eqref{eq:buoyancy} in \eqref{eq:phi_evol}). 
Introducing the vertical buoyancy flux $F_b(\boldsymbol{r},t) = u_z(\boldsymbol{r},t) b(\boldsymbol{r},t)$, one can write the equation for the plane-averaged buoyancy, in the absence of molecular diffusivity, as 
:
\begin{equation}
    \partial_t \langle b \rangle = -\partial_z \langle F_b \rangle = -\frac{\partial \langle F_b \rangle}{\partial \langle \partial_z b \rangle}\, \partial_{zz}  \langle b \rangle\ ,
    \label{eq:flux-grad-buoyancy}
\end{equation}
where the r.h.s is obtained from the chain rule assuming, following Posmentier, that the flux depends only on the vertical buoyancy gradient. 
In the literature the buoyancy evolution is also expressed as a nonlinear diffusion equation of the form: 
\begin{equation}
    \partial_t \langle b \rangle = -\partial_z \langle \mathcal{D}\, \partial_z b \rangle
\end{equation}
where $\mathcal{D}=\mathcal{D}(z,t)$ is a nontrivial function of height.
From \eqref{eq:flux-grad-buoyancy} one notices that the sign of the effective diffusion coefficient $\mathcal{K}(z,t) \equiv -\partial \langle F_b \rangle / \partial \langle \partial_z b \rangle$ discriminates between diffusive and anti-diffusive behavior: in the former, where $\mathcal{K}>0$, small fluctuations in the buoyancy profile are suppressed while in the anti-diffusive regime with $\mathcal{K}<0$ there is a fast growth of small instabilities. 

The relation between vertical flux and vertical gradient of the buoyancy is key to determine whether the formation of density steps is caused by this anti-diffusive mechanism. 
Specifically, if the flux is observed to decrease as the buoyancy gradient gets larger (in absolute value), then perturbations of the linear buoyancy profile get amplified. 
\begin{figure}
    \centering
    \includegraphics[width=\linewidth]{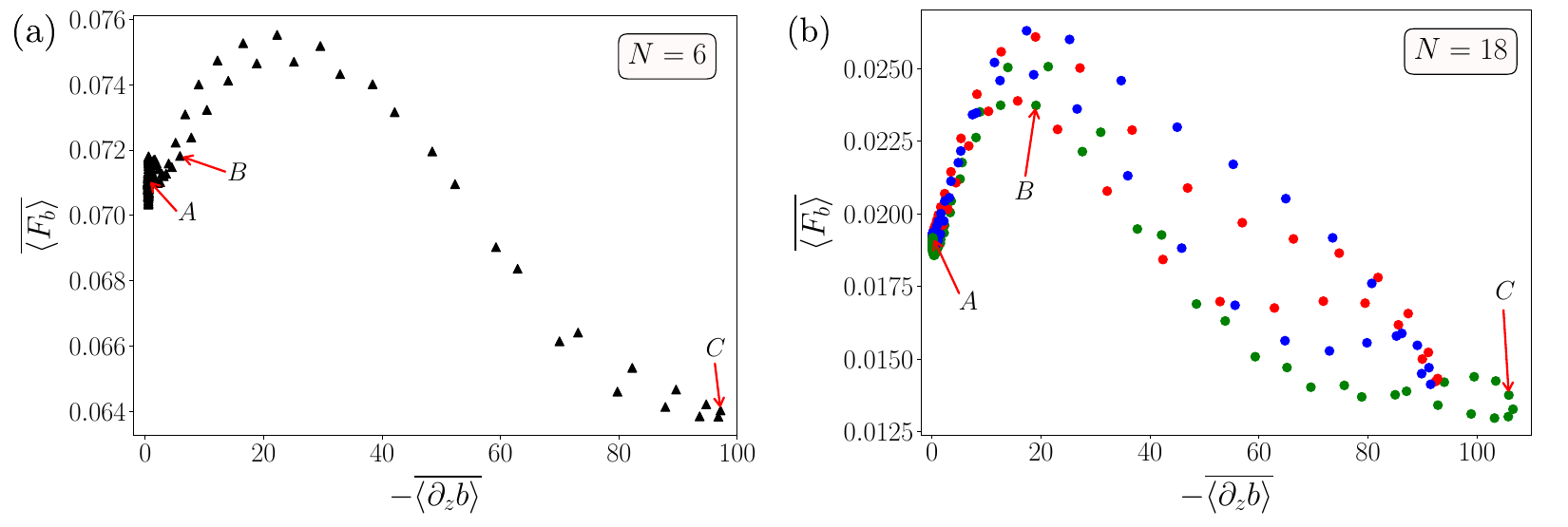}
    \caption{Vertical buoyancy flux as a function of (minus) the vertical buoyancy gradient, with plane- and time-averages performed on both quantities, for (a) $N=6$ and (b) $N=18$. 
    In the former the data are from the final, one-interface state, in the latter from the three-interfaces stationary state. 
    Points $A$, $B$ and $C$ in both panels indicate points of interest along $z$ (respectively: center of layer, interface-layer border, center of interface) as reported in figs.~\ref{fig:vertical_profiles}(a-b).
    In panel (b) points measured from different interfaces -- and their adjacent half-layers -- are shown with different colors (see fig.~\ref{fig:vertical_profiles}(b)).}
    \label{fig:flux-gradient}
\end{figure}
In figure~\ref{fig:flux-gradient} we show the non-monotonic relation found between the vertical buoyancy flux and the vertical buoyancy gradient, both plane-averaged and time-averaged at stationarity, for $N=6$ (panel a) and $N=18$ (panel b).  
We sampled these quantities from the final state with one interface for $N=6$, and from the intermediate state with three interfaces for $N=18$. 
The shape of the curves manifestly resembles the flux curve proposed by \citet{posmentier1977} and verified by experiments \citep{linden1979mixing}. 
As the buoyancy gradient grows in modulus the flux is observed to increase, reach its maximum and then decrease to its minimum. 
Therefore, the portions of stratified fluid having vertical buoyancy gradient larger in magnitude than a critical value (i.e. where the flux has its maximum) will be interested by the anti-diffusive mechanism, that amplifies deviations from the linear density profile. 

Let us point out some features of the flux-gradient relation obtained. 
All points deriving from measurements in the layers (see label $A$, reported also in fig.~\ref{fig:vertical_profiles}) are clumped on the left since homogeneous density implies $\partial_z b = 0$. 
Considering the layer-interface borders as the heights where $\partial_z \phi \equiv \partial_z b + N = 0$ (see label $B$ and fig.~\ref{fig:vertical_profiles}), then points in figure~\ref{fig:flux-gradient} with $-\overline{\langle \partial_z b \rangle} > N$ are measured inside interfaces, therefore the peak and the decreasing part of the flux take place inside interfaces. 
Finally, points with the most negative buoyancy gradient, namely those on the rightmost part of the figures (label $C$), show the lowest vertical flux.
The dispersion observed in the decreasing part of the measurements at $N=18$, much broader than that at $N=6$, is due to appreciable asymmetries existing along $z$ between flux and gradient of buoyancy. 
The one-interface state at $N=6$ show much more ``periodic'' density and flux profiles than the three-interfaces state of $N=18$ (see again fig.~\ref{fig:vertical_profiles} for the $\phi$ profiles in the two cases). 
Furthermore the flux is a highly-fluctuating quantity, that has to be tamed with long-enough time-averages to reduce its dispersion. 
Notably, we see from figure~\ref{fig:flux-gradient} that the (averaged) kinetic-to-potential energy exchange term $\overline{\langle N \phi u_z \rangle} \equiv N \overline{\langle F_b \rangle}$ as a function of the buoyancy gradient show a nice superposition in the two cases. 
This is consistent with experimental data \citep{linden1979mixing} showing that the flux Richardson number $R_f$ -- defined in \cite{park1994turbulent} as $R_f \approx NF_b/\varepsilon_K$ -- plotted as a function of the buoyancy gradient for different experiments, shows a good universal behavior.

Eq.~\eqref{eq:flux-grad-buoyancy} says that stationarity can be reached only locally where $\mathcal{K}=0$ (i.e. at the peak of $\overline{\langle F_b \rangle}$) or where the vertical buoyancy profile changes concavity. 
It is then the omitted dissipative term which stops the unbounded growth of small instabilities and guarantees (meta)stability during the buoyancy evolution. 
\citet{paparella2012clustering}, studying fingering in double-diffusion, considered also the dissipative terms inside the definition of effective flux, finding the similar non-monotonic behavior as a function of the vertical buoyancy gradient.


\section{Decaying vs. merging of interfaces}
\label{sec:decay_merge}

As discovered by \citet{park1994turbulent}, there are two mechanisms responsible for the layer-thickening process (or, equivalently, to the depletion of the high-gradient density interfaces) observed in experiments. 
The first, observed in figs.~\ref{fig:visual_layers}(c)-(e), consists in a ``static'' decay of a weaker interface, that occurs with little to no vertical movement of the high-gradient regions and during which the two neighboring layers become a single, thicker layer. 
The second one, instead, sees pairs of interfaces drifting vertically towards each other until they merge, and the layer enclosed between them disappears. 
\citet{park1994turbulent} noted that both mechanisms take place when $Fr$ is low enough, and \citet{radko2007mechanics} determined analytical conditions of the buoyancy flux for obtaining either a decay or a merging of interfaces. 

\begin{figure}
    \centering
    \includegraphics[width=\linewidth]{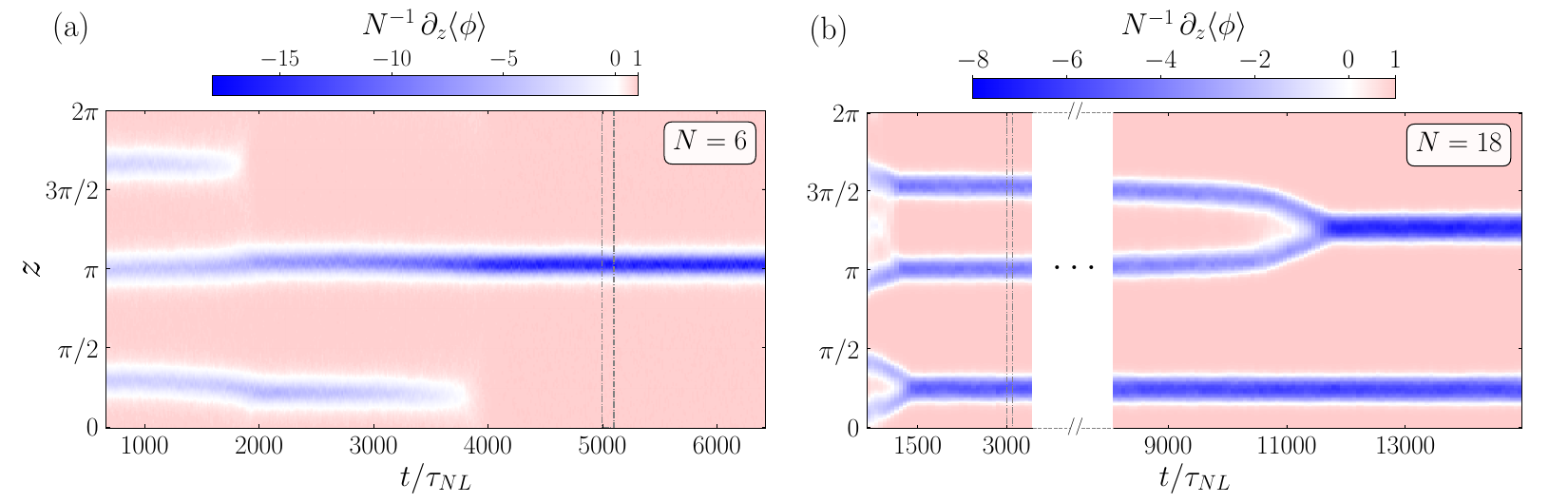}
    \caption{Evolution of the vertical gradient of the plane-averaged density fluctuation $\langle \phi \rangle$, made non-dimensional rescaling with the Brunt-Vais\"al\"a frequency. Large negative values (in blue) correspond to density interfaces, slightly-positive values (in red) to layers. The run with $N=6$ in (a) shows decays of interfaces, while in panel (b) we observe merging of interfaces when $N=18$. The thin strips delimited by dash-dotted lines indicate where the time-averaging for fig.~\ref{fig:vertical_profiles} is computed.}
    \label{fig:decay_merge}
\end{figure}
Figure~\ref{fig:decay_merge} shows a comparison of the interfaces evolution for the cases $N=6$ and $N=18$. 
Colormaps of the (rescaled) vertical gradient of the plane-averaged density fluctuation $\langle \phi \rangle$ reveal that when $N=6$, in panel (a), the weaker interfaces tend to decay with little vertical drift until a single interface dominates, whereas in panel (b) pairs of interfaces come closer and merge together when the stratification is stronger ($N=18$).
While decay and merging events increase the strength (steepness) of the remaining interfaces, the layers keep their stationary value $\partial_z \langle\phi\rangle \sim N$, a relation that can be derived averaging \eqref{eq:phi_evol} on planes neglecting diffusivity, and corresponding to $\partial_z \langle b \rangle \sim 0$. 
Indeed the maximum value assumed by the rescaled vertical gradients $N^{-1} \partial_z \langle \phi \rangle = 1$ corresponds to the limit of gravitational stability, namely a constant density profile.

It is expected that the two interfaces remaining at the final time for $N=18$ will perform a final merging event, but the simulation was not extended further enough to witness it since we estimated, from the growth rate of low-wave-number modes, that it would have taken $\sim 10^4$ turnover times to occur. 
We also remark that the drifting and fusion of interfaces occurring in fig.~\ref{fig:decay_merge}(b) at $t \sim 2000\, \tau_{NL}$ and $t \sim 17000\, \tau_{NL}$ appear to be mediated by the kinetic energy, at variance with the $N=6$ case in which the kinetic energy is almost unaffected by interface decays. 

We also observed an interesting difference regarding the initial generation of the layered state in the two cases. 
While the $N=6$ run shows a sudden transition from linear density profile to a three-interfaces state, for $N=18$ we found the formation of numerous thin layers that undergo many simultaneous (and relatively fast) merging processes until a state similar to the initial one of fig.~\ref{fig:decay_merge}(b) is reached. 
This is in agreement with \citet{park1994turbulent}, where they showed that the layer width grows linearly with the buoyancy scale $U_{rms}/N$. Understanding whether the forcing scale also play a role in determining the shape of the initial step-like profile is left for future studies.


\section{Extreme events and intermittency} 
\label{sec:intermittency}

\begin{figure}
    \centering
    \includegraphics[width=\linewidth]{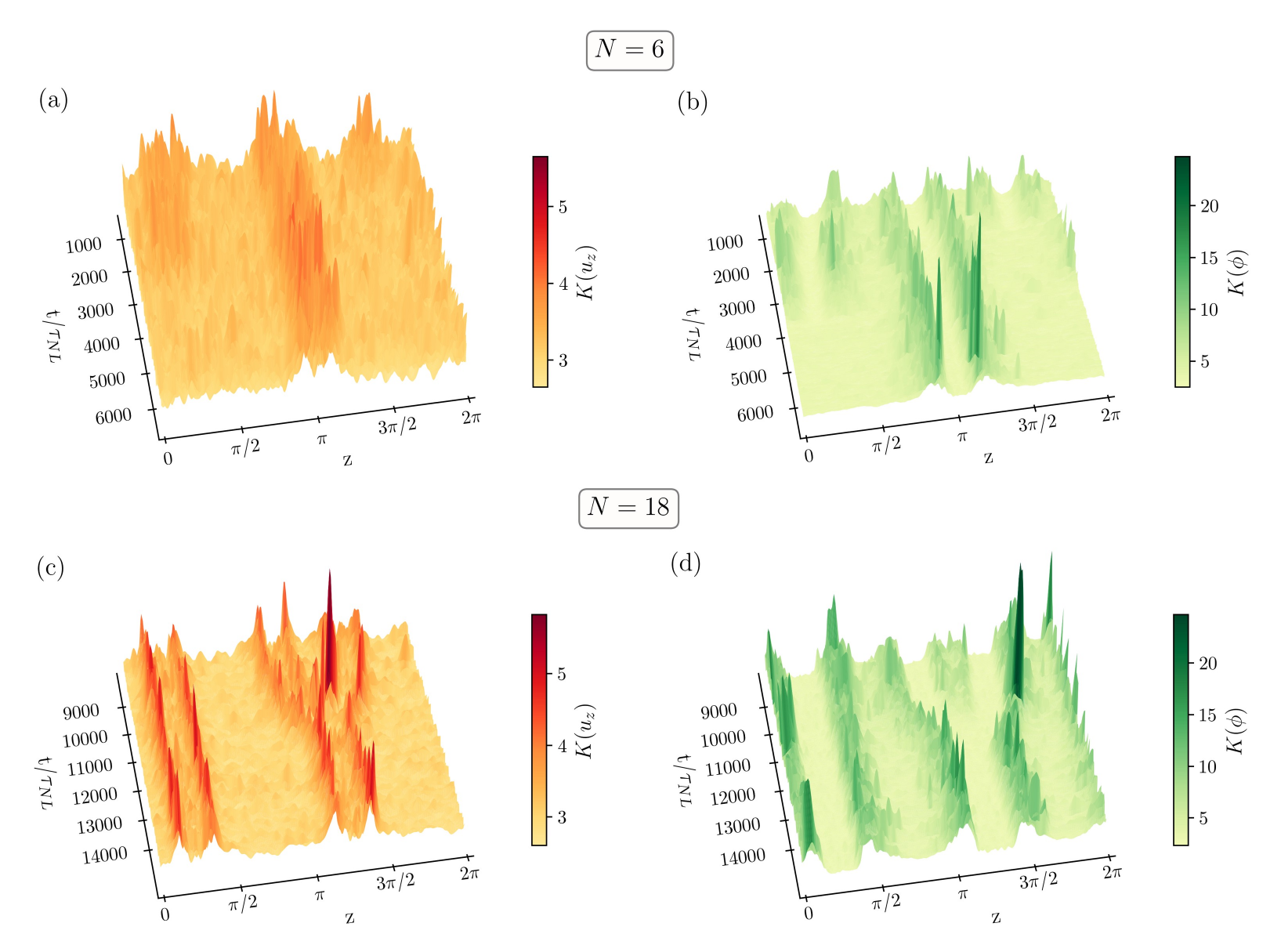}
    \caption{Vertical and time variations of the plane-wise kurtosis $K$, computed for vertical velocity and density fluctuation. Top row shows surface plots, aided by color coding, for the $N=6$ case, bottom row for $N=18$. The flatness factors of vertical velocity are depicted in shades of red in panels (a) and (c); those of density fluctuations, in shades of green, are found in panels (b) and (d).}
    \label{fig:intermittency}
\end{figure}

Despite their reduced capability to develop consistent vertical fluxes due to gravitational stability, oceanic flows manifest, to some extent, a good degree of mixing across large vertical gradients. 
As an example, vertical transports of heat,
salt and nutrients across the main ocean pycnocline are commonly observed, and are indeed crucial for biological activities and the ocean circulation \citep{fernando1991mixing}. 
A significant aspect is that mixing and turbulence at the interfaces, induced by internal wave breaking and shear instabilities, feature extreme and sporadic events. 
Recent numerical works \citep{rorai2014bursts, feraco2018vertical, marino2022turbulence} have highlighted the occurrence of strong intermittent vertical drafts, dynamically related to notable enhancement of local density fluctuations, signifying that vertical transport of mass and momentum is an abrupt event in space and time. 
\citet{feraco2018vertical} showed that regions in the spatial domain where the maximum kurtosis of $u_z$ and $\phi$ are observed presented the typical flattened structure. 
However, the time extension of their numerical simulations -- approximately $25$ large-scale turnover times -- places their results before the complete development of the layer-interface structure. 
It is instead interesting to investigate the fully-formed layered structure to study the statistical properties, especially intermittent events, inside layers and interfaces. 
To this end we compute the plane-wise kurtosis (or flatness factor):
\begin{equation}
    K(z,t; f) = \langle (f-\langle f \rangle)^4 \rangle\ /\ \langle (f-\langle f \rangle)^2 \rangle^2\ ,
    \label{eq:kurt_xy}
\end{equation}
with $f=f(x,y,z,t)$ a generic field component,
of both vertical velocity $u_z$ and rescaled density fluctuation $\phi$. 
Figure~\ref{fig:intermittency} displays the evolution, in time and along the vertical direction, of the flatness factors of vertical velocity (in red) and the density fluctuation (in green) for the two layered cases.
With the aid of fig.~\ref{fig:decay_merge}, a general observation is that there are very distinct statistical behaviors in layers and interfaces, and indeed the locations of the interfaces, as well as the coarsening processes, are identified from large-kurtosis events. 
In all plots the typical Gaussian value $K=3$ is observed inside layers where the density is quasi-homogeneous. 
Instead in proximity to layer-interface borders, namely where local maxima and minima of $\langle\phi\rangle$ are attained, the flatness factors show sharp and localized increases (for instance, in fig.~\ref{fig:intermittency}(d) we measure $\text{max}_{z,t}[K(z,t;\phi)]/\text{min}_{z,t}[K(z,t;\phi)]>17$). 
Unexpectedly, in the interior of the density interfaces the kurtosis drops close to the Gaussian value.

It appears that, in general, very strong values of the flatness factors are localized where the vertical density profile shows the largest curvature, transitioning from being nearly flat at the layers to extremely steep in the interfaces. 

\begin{figure}
    \centering
    \begin{subfigure}{0.49\textwidth}
         \centering
         \includegraphics[width=\linewidth]{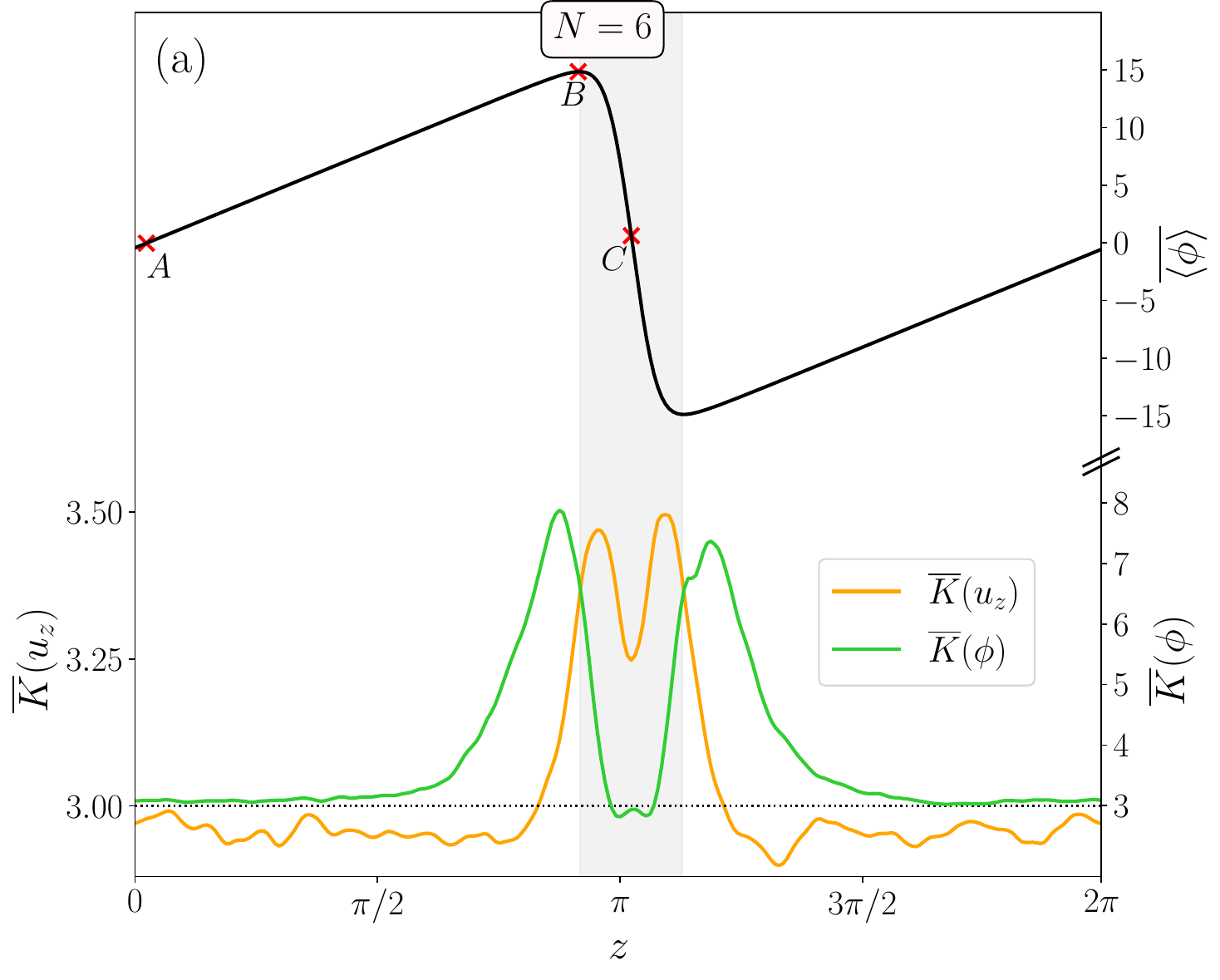}
    \end{subfigure}
    \begin{subfigure}{0.49\textwidth}
         \centering
         \includegraphics[width=\linewidth]{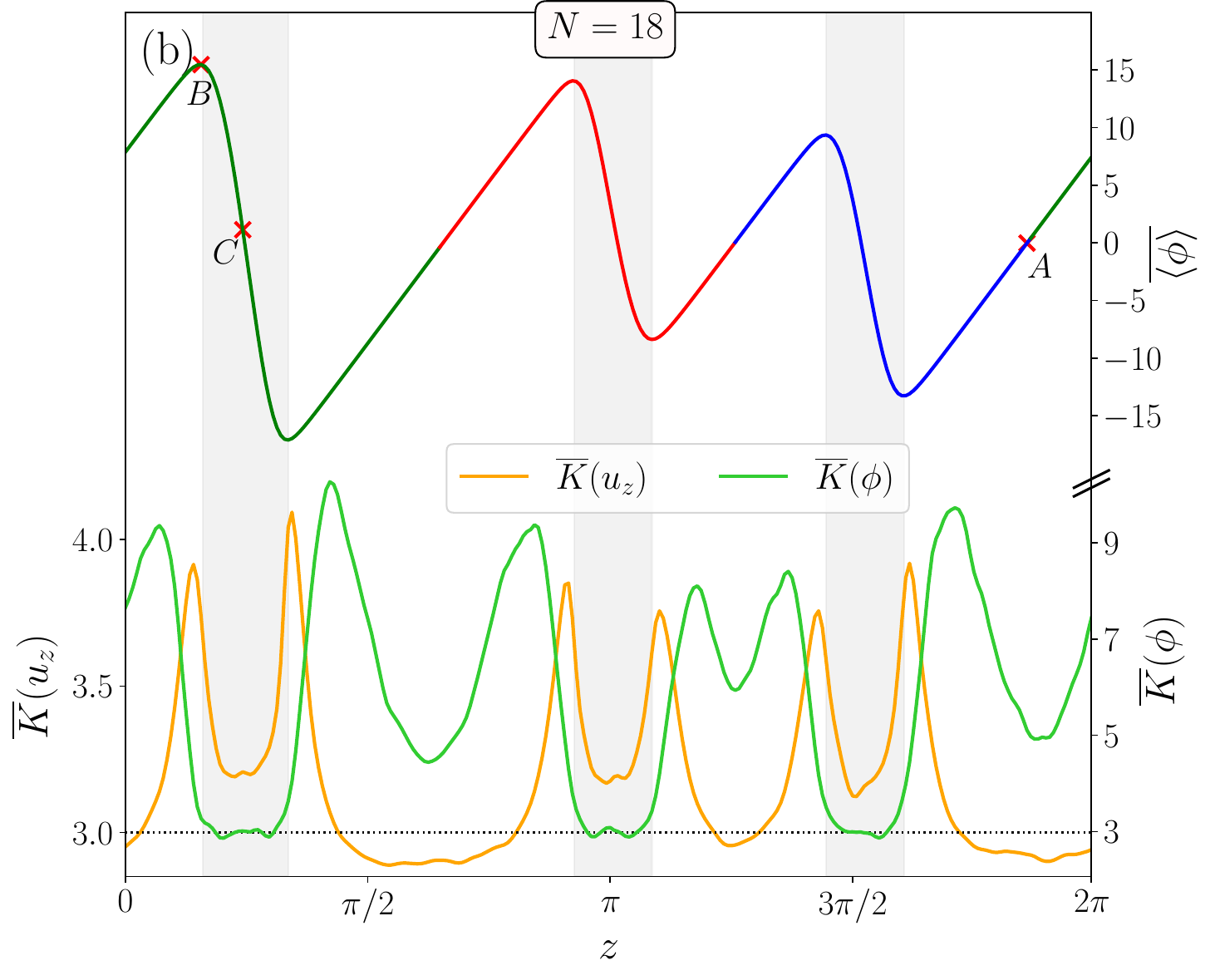}
    \end{subfigure}
    \caption{Plane- and time-averaged vertical profiles of density fluctuation field (top plots) and kurtosis of vertical velocity and scalar (bottom plots) for (a) $N=6$ and (b) $N=18$. 
    Time averages are taken in the intervals $t/\tau_{NL} \in [5000,5100]$ for panel (a) and $t/\tau_{NL} \in [3000,3100]$ for panel (b): these ranges are highlighted by dash-dotted lines in fig.~\ref{fig:decay_merge}. 
    Shaded grey areas indicate interfaces extensions, as delimited by local extrema of $\overline{\langle \phi \rangle}$. 
    Points $A$, $B$ and $C$ indicate points of interest (center of layer, interface-layer border, center of interface respectively) that are also reported in fig.~\ref{fig:flux-gradient}.
    In the top plot of (b) each interface, with its adjacent half-layer, is shown with the same color as fig.~\ref{fig:flux-gradient}(b). 
}
    \label{fig:vertical_profiles}
\end{figure}
A more thorough inspection reveals that peaks of $K(u_z)$ are in correspondence to interface-layer borders (considering these as the vertical coordinates where $\langle \phi \rangle$ has local extrema), while peaks of $K(\phi)$ always ``enclose" those of the velocity kurtosis, and are positioned towards the bulk of the adjacent layers even if very close to interface-layer borders. 
This is shown in fig.~\ref{fig:vertical_profiles}(a) for $N=6$ and \ref{fig:vertical_profiles}(b) for $N=18$, where we compare the typical vertical profile of $\phi$ with the vertical profiles of both flatness factors. 
To smooth out temporal fluctuations in the fourth-order moments all profiles have been time-averages over $\sim 50$ consecutive snapshots.

We conclude by remarking that intermittent events are strongly linked to the full formation of density layers and interfaces: for this reason it is not to be excluded that \citet{feraco2018vertical} did not measure large kurtosis of the vertical velocity below a critical $Fr$ value because inter-event times become much large as $N$ increases. 
Indeed we observe that for our strongest-stratified case ($Fr=0.076$) the volume-averaged kurtosis of $u_z$ grows noticeably in the build-up of strong merging events, and decays after merging has taken place (not shown). 
If we suppose merging processes to be hindered by stronger buoyancy forces, then decreasing $Fr$ would make intermittent bursts very sporadic. 
The verification of such assumption is left for future studies.


\section{Conclusions and future perspectives} 
\label{sec:conclusion}

We have presented a numerical study on forced Boussinesq flows with unprecedented total time integration (up to $\sim 2 \times 10^4$ large-scale turnover times) to investigate the formation and late-time evolution of stacked density layers and interfaces, starting from a linear vertical density profile. 
In an attempt to mimic experimental conditions and hinder the unbounded growth of shear modes a damping term on the large-scale velocity is added, rarely considered in simulations of non-rotating stratified flows.
The focus of the present work has been to highlight the differences in overall long-time evolution at varying the stratification strength. 
Concerning the energy content and density redistribution of the flow, the simulation with smaller Brunt-Vais\"al\"a frequency, $N=3$, displays no layering, and kinetic and potential energy reach stationary values as HIT does; 
doubling the frequency to $N=6$ one observe a stationary kinetic energy but the emergence of density layers and interfaces, the latter undergoing several decay processes associated with increases of potential energy; 
raising the frequency to $N=18$ still yields a layered fluid, yet the dynamical mechanism of interfaces coarsening switches from decay to merging of pairs of interfaces, and this phenomenological change is also reflected by the kinetic energy displaying surges in unison with the merging events. 
The transitions between these behaviors is consistent with the experimental results of \citet{park1994turbulent}, who observed analogous features in the density interfaces when increasing the Richardson number -- a parameter quantifying the buoyancy-to-shear forces ratio and thereby growing as $N$ grows. 
We observed how layer formation and evolution is equivalent to a progressive accumulation of potential energy to the lowest non-zero modes, and verified that the instability mechanism that triggers the creation of a staircase profile (and its pseudo-stationarity) is well captured by the Phillips-Posmentier picture.
Intermittent features in the statistics of vertical velocity and density fluctuation were sought: while inside layers both variables are Gaussian-distributed, which is expected since in those subdomains the flow differs only slightly from HIT, strong and rare events are found where layers and interfaces meet, namely where the transition from an almost flat to a very steep local density profile takes place. 

The slowness of the initial layering formation, and the much slower dynamics of interface coarsening -- tens of thousands of turnover times might pass between two of such events -- justifies the little emphasis on these phenomena found in the literature from a computational perspective. 
It is hard at the moment to make estimations about typical inter-event times and relations with the stratification strength, considering that only two layered runs are available and especially that the two coarsening processes shown are totally different.

This work opens the way to further numerical studies, aimed both at exploring other areas of the parameter space and at uncovering other aspects of the fully-layered state, such as the anisotropic spectral transfers among flow components and among linear modes, the dependence of the initial layered state on the external forcing, or the mechanism leading to intermittent bursts inside interfaces (and generally in stratified flows).\\


\backsection[Acknowledgements]{
We are greatly indebted to Gabriella Schirinzi for developing the numerical code for the Boussinesq equations. 
A.S.L. thanks Centrale M\`editerran\'ee and CNRS-IRPHE (Marseille, France) for the kind hospitality, where this worked was discussed.
}

\backsection[Funding]{
We acknowledge EuroHPC Regular Access, project No. EHPC-REG-2021R0049, for the computational resources.
This work was supported by the European Research Council (ERC) under the European Union’s Horizon 2020 research and innovation programme Smart-TURB (Grant Agreement No. 882340). 
N.C. acknowledges support from the National Recovery and Resilience Plan (PNRR), Mission 4 Component 2 Investment 1.1 - Call No. 104 – Project: CO-SEARCH, code 202249Z89M – (CUP E53D23001610006), 
and from Fondo Italiano per la Scienza 2022-2023 (FIS2), Project title {\it Data-driven and equation- based tools for Deep understanding of multi-scale Complex Turbulent Flows}, CUP E53C24003760001. 
A.S.L. acknowledges financial support under the National Recovery and Resilience Plan (PNRR), by the Italian Ministry of University and Research (MUR), funded by the European Union — NextGenerationEU — Project Title {\it Numerical and Experimental Investigation of Small Plastics Breakup in Complex Flows (BREAKUP)}, Contract P20225AEF4 — CUP B53D23028750001. 
}

\backsection[Declaration of interests]{
The authors report no conflict of interest.
}





\bibliographystyle{jfm}
\bibliography{bib}

\end{document}